\documentclass[twocolumn,superscriptaddress,nofootinbib,showkeys]{revtex4-1}
\usepackage[utf8]{inputenc}
\usepackage{amsmath,amssymb}
\usepackage{bm,cancel}
\usepackage{graphicx}
\usepackage[dvipsnames]{xcolor}
\graphicspath{{figures/}}

\newcommand{\ba}{\bar{a}}
\newcommand{\bc}{\bar{c}}
\newcommand{\bJ}{\bar{J}}
\newcommand{\blambda}{\bar{\lambda}}

\newcommand{\linking}{\mathcal{L}_k}
\newcommand{\linkingself}{\mathcal{L}_S}
\newcommand{\linkingfrenet}{\linking,_{\text{Frenet}}}
\newcommand{\twist}{\mathcal{T}_w}
\newcommand{\twistfrenet}{\twist,_{\text{Frenet}}}
\newcommand{\writhe}{\mathcal{W}_r}
\newcommand{\writhefrenet}{\writhe,_{\text{Frenet}}}
\newcommand{\sn}{\text{sn}}
\newcommand{\sgn}{\text{sgn}}

\newcommand{\amend}[1]{#1}
\newcommand{\Calugareanu}{Călugăreanu }

\begin{document}
%
\title{Non-planar elasticae as optimal curves for the magnetic axis of stellarators}%
\author{D. Pfefferlé}%
\affiliation{The University of Western Australia, 35 Stirling Highway,
  Crawley WA 6009, Australia} \author{L. Gunderson}
\affiliation{Princeton Plasma Physics Laboratory, Princeton, New
  Jersey 08543, USA} \author{S.R. Hudson} \affiliation{Princeton
  Plasma Physics Laboratory, Princeton, New Jersey 08543, USA}
\author{L. Noakes} \affiliation{The University of Western Australia,
  35 Stirling Highway, Crawley WA 6009, Australia}
\date{\today}

\begin{abstract}
  \amend{The problem of finding an optimal curve for the target magnetic axis
  of a stellarator is addressed}. Euler-Lagrange equations are derived
  for finite length three-dimensional curves that extremise their
  bending energy while yielding fixed integrated torsion. The obvious
  translational and rotational symmetry is exploited to express
  solutions in a preferred cylindrical coordinate system in terms of
  elliptic Jacobi functions. These solution curves, which, up to
  similarity transformations, depend on three dimensionless
  parameters, do not necessarily close. Two closure conditions are
  obtained for the vertical and toroidal displacement (the radial
  coordinate being trivially periodic) to yield a countably infinite
  set of one-parameter families of closed non-planar curves. The
  behaviour of the integrated torsion (Twist of the Frenet frame), the
  Linking of the Frenet frame and the Writhe of the solution curves is
  studied in light of the \Calugareanu theorem. A refreshed
  interpretation of Mercier's formula for the on-axis rotational
  transform of stellarator magnetic field-lines is proposed.
\end{abstract}


\maketitle

\section{Introduction}
In toroidal fusion devices such as the tokamak and the stellarator,
high rotational transform - the average linking of neighbouring
field-lines per revolution around the axis - is essential for plasma
confinement~\cite{spitzer-1958}. In tokamaks, which are axisymmetric
devices, the winding of the magnetic field is obtained by varying the
poloidal flux in time, thereby inducing strong toroidal (parallel)
currents (or curl of the inner magnetic field). The duration of a
tokamak discharge is thus limited to the duration of the ramp-up and
ramp-down of the central solenoid stack, \amend{although there are
  ways to sustain current through radio waves (cyclotron resonance)
  \cite{sauter-2000} or by exploiting kinetic effects (bootstrap)
  \cite{kikuchi-azumi}}. In stellarators, the confining field is
static and net toroidal currents are avoided. The external
current-carrying coils are arranged in such a way that the vacuum
magnetic field produced wraps tightly around the plasma, thereby
linking the magnetic axis by the helical trajectory of its
field-lines. \amend{The main reason the production cost of a
  stellarator is so high is because the shape of its coils can be
  extremely complex. Optimising the coil geometry is an important
  component of stellarator design \cite{landreman-2017} and a
  prerequisite for the economic viability of the stellarator concept.}

\amend{The stellarator optimisation problem is extremely difficult in
  that the number of degrees of freedom is far too large to be
  explored empirically; design points must be chosen through
  well-organised computations \cite{gates-2017}. A potential caveat is
  that the choice of coordinates and magnetic field representation
  will favour certain classes of solutions. For example, a Fourier
  expansion in terms of the toroidal angle will not allow for plasma
  shapes with vertical or radial portions, although it is entirely
  possible that such configurations can result in simpler coils and
  better confinement performances. In this sense, numerical
  computations must be guided by the theoretical understanding of
  simple models, that can serve as reliable benchmark cases for code
  verification and validation.}

It is common knowledge that high rotational transform at the magnetic
axis of stellarators can be achieved without inducing current by means
of two independent effects~\cite{mercier-1964,helander-2012}: i)
rotating an elliptic plasma boundary along a circular (planar) axis,
as for example in the Large Helical Device (LHD) in Toki,
Japan~\cite{lhd}, ii) harnessing finite integrated torsion from a
non-planar magnetic axis, as for example in the Wendelstein 7-X in
Greifswald, Germany~\cite{w7x}.
In the first case, penetration of the helical component is expected to
decay as $r^{m-1}$, where $r$ is the minor radius of the torus and $m$
is the poloidal mode number of the applied boundary rotation (for an
elliptic boundary $m=2$). The plasma edge thus receives more shear
than the axis. In the second case, the rotational transform at the
magnetic axis arises purely from its geometry. It can be imposed, up
to a certain extent, almost independently from the rotation of the
boundary. \amend{A recent coil design exercise using the FOCUS code
  \cite{focus} has shown that increasing ellipticity of the boundary
  tends to increase the coil complexity more so than increasing the
  integrated torsion \cite{hudson-2018}.}

It is thus legitimate to ask what smooth curve provides maximum
integrated torsion without being too ``complex''. More precisely: What
are the closed curves of fixed length and given integrated torsion
that minimise bending energy
? 
This particular question is addressed in a series of mathematical
papers~\cite{langer-singer-1984,langer-singer-1996,ivey-singer-1999}. The
analysis therein is \amend{repeated and extended with a focus on being able to 
  control the integrated torsion of the magnetic axis and produce a
  maximum amount of rotational transform. The representation of
  solution curves is exploited to categorise families of ideal target
  magnetic axes and starting points for efficient stellarator
  designs. Our derivation deviates from~\cite{ivey-singer-1999} in the
  fact that the length of the curve is held fixed and the maximum
  curvature becomes a dependent variable, whereas the opposite
  assumption is made in their work.}

The strategy is to first obtain the solution to the variational
problem of minimising the total squared curvature of a curve with
fixed length and prescribed integrated torsion. The solution curves do
not necessarily close on themselves, therefore additional conditions
are derived to constrain the set of free parameters. The result is a
countably infinite set of one-parameter families of closed curves
isotopic to torus knots (including the unknot). It is then highlighted
that smooth deformations of the solution curves near states of
inflexion give rise to severe discontinuities in the integrated
torsion, which on physical grounds cannot affect the on-axis
rotational transform. This inconsistency is resolved by applying the
\Calugareanu theorem to the linking of magnetic field-lines separately
to the linking of the Frenet frame on the magnetic axis. In fact, it
is shown that the relevant quantity for the on-axis rotational
transform due to non-planar geometry is the so-called \emph{Writhe} of
the supporting curve. A quantitative estimate of the Writhe (and
consequently the on-axis rotational transform) is derived for our
closed energy-minimising (elastic) curves. The well-known formula by
Mercier \cite{mercier-1964} is revisited to provide a clear
interpretation of its mysterious integer of topological
origin.

After some definitions, section \ref{sec:definitions} recalls the
properties of the Frenet frame for three-dimensional curves. Section
\ref{sec:euler-lagrange} formulates the variational problem of
minimising bending energy subject to constraints on length and
integrated torsion, and derives the associated Euler-Lagrange
equations. Section \ref{sec:noether} exploits the symmetries of the
Lagrangian in order to yield two Noether fields. Section
\ref{sec:analytic} derives analytic solutions to the Euler-Lagrange
equations in a special cylindrical coordinate system. Section
\ref{sec:closures} lists a series of conditions for the solution
curves to be periodic and closed. Section \ref{sec:link-twist-writhe}
assesses the integrated torsion and expresses the linking of the
Frenet frame as well as the Writhe of closed solution curves based on
the \Calugareanu theorem. The effect of non-planar geometry on the
on-axis rotational transform is quantified for $n$-periodic unknotted
closed solution curves and the conjugate $(n-1,n)$-torus
knots. Section \ref{sec:conclusion} concludes the work.

\section{Definitions}
\label{sec:definitions}
\amend{The target magnetic axis of a stellarator} is represented by a smooth
curve $\bm{x}(s)$ (of class $C^\infty$), parametrised by its
arc-length $s\in [0,L]$, where $L$ is its total length. This
representation is such that the velocity vector has unit length,
\begin{align}
  \label{eq:natural_param}
  \bm{x}'\cdot\bm{x}' = 1,
\end{align}
where $\bm{x}'=d\bm{x}/ds$. One can verify that there is no loss of
generality in choosing the arc-length parametrisation, only for an
explicit constraint (Lagrange multiplier function) appearing in the
variational formulation; the same Euler-Lagrange equations are
obtained starting with an arbitrary parametrisation where only the
length of the curve is constrained~\cite{langer-singer-1996}. The
arc-length parametrisation significantly reduces the algebra.

The fundamental theorem of curves states that every regular curve in
$\mathbb{R}^3$ with non-zero curvature is completely determined up to
isometries by its curvature and torsion functions\amend{, respectively $\kappa$ and $\tau$}, via the Frenet-Serret formulae,
\begin{align}
  \label{eq:frenet-serret}
  \begin{pmatrix}
    \bm{x}\\
    \bm{t}\\
    \bm{n}\\
    \bm{b}
  \end{pmatrix}'
  &=\begin{pmatrix}
    0 & 1 & 0 & 0 \\
    0 & 0 & \kappa & 0\\
    0 & -\kappa & 0 & \tau \\
    0 & 0 & -\tau & 0
    \end{pmatrix}
 \begin{pmatrix}
 \bm{x}\\
 \bm{t}\\
 \bm{n}\\
 \bm{b}
  \end{pmatrix},
\end{align}
where $\{\bm{t}, \bm{n}, \bm{b}\}$ - the tangent, normal, and bi-normal vector triad - forms the orthonormal Frenet frame,
\begin{equation}
\begin{array}{rl}
\bm{x}' & \overset{(\ref{eq:natural_param})}{=}\bm{t} \\
  \bm{x}'' & = \bm{t}' = \kappa \bm{n} \\
  \bm{x}''' &= -\kappa^2 \bm{t} + \kappa'\bm{n} + \kappa \tau \bm{b} \\
  \bm{x}'''' & = -3\kappa'\kappa \bm{t} + (\kappa''-\kappa^3-\kappa\tau^2)\bm{n} + (2\kappa'\tau + \kappa \tau')\bm{b}
\end{array}
\label{eq:derivatives}
\end{equation}
where $\bm{b} = \bm{t}\times\bm{n}$.

The curvature and torsion of the curve correspond to
\begin{align}
  \kappa^2(\bm{x}'') &= \bm{x}''\cdot\bm{x}'' \overset{(\ref{eq:natural_param})}{=} -\bm{x}'\cdot\bm{x}''',\label{eq:curvature}\\
 \tau(\bm{x}',\bm{x}'',\bm{x}''') &= \frac{(\bm{x}'\times\bm{x}'')\cdot\bm{x}'''}{\kappa^2}\label{eq:torsion}.
\end{align}
These functions are noted to scale with the inverse of the curve's
length, i.e.  $\kappa\propto \tau \propto L^{-1}$. This fact is used
later to adimensionalise the parameter space of the solution
curves. The following identities are also listed for later use:
\begin{align}
  \label{eq:tau_derivatives}
  \frac{\partial \tau}{\partial \bm{x}'} &=\tau \bm{t} + \kappa \bm{b}, &
    \left(\frac{\partial \tau}{\partial \bm{x}'}\right)'&= \tau'\bm{t} + \kappa'\bm{b},\\
  \frac{\partial \tau}{\partial \bm{x}''} &= -\frac{\tau}{\kappa}\bm{n} - \frac{\kappa'}{\kappa^2}\bm{b}, &
  \frac{\partial \tau}{\partial \bm{x}'''}  &= \frac{1}{\kappa}\bm{b},\\
  \frac{\partial \tau}{\partial \bm{x}''} &= \left(\frac{\partial \tau}{\partial \bm{x}'''}\right)'.
 & \label{eq:cancellation}
\end{align}

\amend{Owing to the classical Euler-Bernoulli model of thin elastic rods, we consider the \emph{bending energy} of a curve to be proportional to the total squared curvature,
\begin{align}
\mathcal{E} := \int_0^L \tfrac{1}{2} \kappa^2 ds = \int_0^L \tfrac{1}{2} \bm{x}''\cdot \bm{x}'' ds,
\end{align}
as a measure of the complexity of the magnetic field and therefore a component of the total cost of the coil design.}

\amend{Critical points of the bending energy functional for given boundary
conditions are well-known curves called \emph{elasticae} and have been
extensively studied in the context of beam deflection.}
\amend{Among all closed curves, Fenchel's theorem \cite{fenchel-1951}
  states that $\mathcal{E}\geq 4\pi^2/L$, where the lower bound is
  achieved by the planar circle - the most trivial closed elastica
  with constant curvature and zero torsion. Interestingly, there is a
  countably infinite set of closed non-circular elasticae
  \cite{langer-singer-1984}, which however correspond to "saddle
  points" of the bending energy functional
  \cite{langer-singer-1985}. A constraint on the non-planarity thus
  needs to be added to the variational problem in order to yield
  stable non-planar optimal stellarator axes. A convenient choice is
  to control the \emph{average torsion}, referred hereafter by the
  following dimensional quantity}
\begin{align}
  \label{eq:torsion_constraint}
  <\tau> := \frac{1}{L}\int_0^L \tau  ds.
\end{align}
We will refer to the \emph{integrated torsion} - or the \emph{Twist of
  the Frenet frame} - by the adimensional quantity
\begin{align}
  \twistfrenet :=  \frac{1}{2\pi}\int_0^L\tau ds = \frac{L}{2\pi}<\tau>.
\end{align}

\section{Euler-Lagrange equations}
\label{sec:euler-lagrange}
The curve of minimal bending energy and prescribed integrated torsion
is the extremal curve of the \amend{objective functional}
\begin{align}
  \label{eq:action}
  S[\bm{x}] =\int_0^L ds\left[ \frac{1}{2}\kappa^2  + \frac{1}{2}\Lambda(\bm{x}'\cdot\bm{x}' -1 ) + \lambda_2 (\tau - \tau_0)\right],
\end{align}
\amend{which consists of three terms: i) the bending energy, ii) a
  term enforcing the arc-length parametrisation of equation
  (\ref{eq:natural_param}) via the Lagrange multiplier function
  $\Lambda(s)$ and iii) a term to control the integrated torsion via
  the scalar Lagrange multiplier $\lambda_2$, so that
  $<\tau>=\tau_0$. The relative sign between the terms is absorbed in
  the Lagrange multipliers. The whole integrand is commonly referred
  to as the \emph{Lagrangian}}.

Variation of the functional $S$ with respect to $\bm{x}$ yields
\begin{multline}
 \delta S = \int_0^L\!ds\Bigg\{ \bm{E}\cdot\delta\bm{x} +
  \frac{d}{ds}\left[\lambda_2\frac{\partial\tau}{\partial
      \bm{x}'''}\cdot\delta\bm{x}'' + \bm{x}''\cdot\delta\bm{x}'\right. \\
  + \left. \left(-\bm{x}''' + \Lambda\bm{x}' + 
      \lambda_2\frac{\partial\tau}{\partial \bm{x}'}\right)\cdot \delta\bm{x}\right]\Bigg\}, 
\end{multline}
where the Euler-Lagrange equations are identified as
\begin{equation}
  \label{eq:euler-lagrange}
\bm{E}(\bm{x}',\bm{x}'',\bm{x}''',\bm{x}'''') =  \bm{x}'''' - (\Lambda \bm{x}')' - \lambda_2  \left(\frac{\partial \tau}{\partial \bm{x}'}\right)' = 0.
\end{equation}
A simplification occurred thanks to (\ref{eq:cancellation}). The
differential vector equation (\ref{eq:euler-lagrange}) can be written
in terms of the curvature and torsion functions in the Frenet frame,
as
\begin{multline}
-(\Lambda' + 3\kappa'\kappa +\lambda_2\tau')\bm{t} + (\kappa''-\kappa^3 - \kappa\tau^2 - \Lambda\kappa)\bm{n} \\
+ [\kappa'(2\tau-\lambda_2) + \kappa\tau']\bm{b} = 0.
\end{multline}
After integrating once, this leads to the following relations:
\begin{align}
   & \Lambda(s) = \lambda_1 -\tfrac{3}{2}\kappa^2 - \lambda_2 \tau,  \\
   & \kappa^2\left(2\tau - \lambda_2\right) = c, \label{eq:torsion_relation}\\
&(\kappa')^2 + \frac{1}{4}(\kappa^2-2\lambda_1)^2 + \kappa^2(\tau-\lambda_2)^2 = J^2, \label{eq:curvature_quadrature}
\end{align}
where $\lambda_1$ is a scalar Lagrange multiplier related to the
constraint on total length, and $c$ and $J^2$ are integration
constants. The four parameters, $\lambda_1$, $\lambda_2$, $c$ and
$J^2$, completely determine the size and shape of the solution to the
Euler-Lagrange equations (up to isometries) \cite{ivey-singer-1999}.

\section{Symmetries and Noether charges}
\label{sec:noether}
The Euler-Lagrange equation (\ref{eq:euler-lagrange}) can be
immediately integrated to yield
$\bm{x}''' - \Lambda\bm{x}' - \lambda_2 \left(\frac{\partial
    \tau}{\partial \bm{x}'}\right) = \bm{J}$, where $\bm{J}$ is the
constant force from the end points propagating along the
curve. Translations and rotations constitute isometries in
$\mathbb{R}^3$ and are obvious symmetries of the Lagrangian. From
Noether's theorem, one learns that $\bm{J}$ represents the conserved
quantity related to the invariance of the Lagrangian under
translations, $\delta \bm{x} = \bm{\Delta}$,
$\delta\bm{x}'=\delta\bm{x}''=0$ where $\bm{\Delta}$ is an arbitrary
constant vector. Thus, the curve is such that the following vector
field is constant:
\begin{align}
  \bm{J} &= \frac{1}{2}(\kappa^2 -2 \lambda_1)\bm{t}
           +\kappa'\bm{n}
           +\kappa(\tau-\lambda_2 )\bm{b}.
\end{align}
Thanks to the invariance of the Lagrangian under rotation,
$\delta\bm{x} = \bm{\Omega}\times\bm{x}$,
$\delta\bm{x}'=\bm{\Omega}\times\bm{x}'$ and
$\delta\bm{x}''=\bm{\Omega}\times\bm{x}''$ where $\bm{\Omega}$ is an
arbitrary constant vector, we obtain another Noether charge $\bm{A}$,
related to the torque from the end points. Therefore, the curve is
such that the following vector field,
\begin{align}
\label{eq:killing}
\bm{I}(s) &= \lambda_2 \bm{t} + \kappa\bm{b}= \bm{A} + \bm{x}\times\bm{J}
\end{align}
generates an isometry (translation along $\bm{A}$ and rotation around
$\bm{J}$). 
%
%
%
We see from (\ref{eq:killing}) that, under translation $\bm{x}\mapsto \bm{x}+\bm{\Delta}$, $\bm{A}\mapsto \bm{A}-\bm{\Delta}\times \bm{J}$. By choosing $\bm{\Delta} =\bm{J}\times \bm{A}/J^2$, the charge $\bm{A}\mapsto \bm{J} (\bm{A}\cdot\bm{J})/J^2$ can be aligned with $\bm{J}$, such that only the only relevant component is
\begin{equation}
\label{eq:link4}
  \bm{A}\cdot\bm{J}= \bm{I}\cdot\bm{J} = \frac{1}{2} c -\lambda_1\lambda_2 := a J^2.
\end{equation}

By virtue of rotational invariance, we mention that the solution
curves respect \emph{stellarator symmetry}~\cite{dewar-1998}, which is
equivalent to a $180-$degree rotation about a characteristic axis in
the $xy$ plane.

Reparametrisation of the curve does not affect its properties. In
particular, the reversal of the arc-length $s\mapsto -s$ together with
$\bm{J}\mapsto-\bm{J}$, $\bm{A}\mapsto-\bm{A}$ and
$\tau_0\mapsto-\tau_0$ represent the same solution. Furthermore, the
Lagrangian is invariant under a parity transformation
$\bm{x}\mapsto-\bm{x}$ together with $\tau_0\mapsto-\tau_0$,
$\lambda_2\mapsto -\lambda_2$, $c\mapsto-c$ and
$\bm{J}\mapsto-\bm{J}$. Solution curves with negative integrated
torsion can be related to the ones with positive via orthogonal
(rotation and parity) and arc-length-reversal
transformations.

\section{Analytic solution}
\label{sec:analytic}
\subsection{Curvature function}
Equations (\ref{eq:torsion_relation}-\ref{eq:curvature_quadrature})
admit the following solution \cite{ivey-singer-1999},
\begin{align}
\kappa^2(s) = \kappa^2_0\left[1-\frac{p^2}{w^2}\sn^2\left(t,p\right)\right],\quad t=\frac{\kappa_0}{2w}(s-s_0),
\label{eq:kappa_solution}
\end{align}
where $\sn(t,p)$ is the Jacobi Elliptic sine function satisfying
$(d\sn/dt)^2=(1-\sn^2)(1-p^2\sn^2)$. The maximum curvature
$\kappa_0>0$ and parameters $0\leq p\leq w\leq 1$ are (non-linearly)
related to the original set through
\begin{align}
4\lambda_1-\lambda_2^2&= \frac{\kappa_0^2}{w^2}(3w^2-p^2-1), \label{eq:link1}\\
  c^2 & = \frac{\kappa_0^6}{w^4}(w^2-p^2)(1-w^2), \label{eq:link2}\\
4J^2 &=(\kappa_0^2-2\lambda_1)^2+\kappa_0^2\left(\frac{c}{\kappa_0^2}-\lambda_2\right)^2.\label{eq:link3}
\end{align}
Since $\kappa_0\propto L^{-1}$, the following adimensional parameters
are used to factor out the uniform scaling of solutions:
\begin{align}
  \bc &= c/\kappa_0^3, & 
  \bJ &=J/\kappa_0^2, &
  \ba &=a \kappa_0, \\
  \blambda_1&=\lambda_1/ \kappa_0^2, &
  \blambda_2 &=\lambda_2/ \kappa_0 .&
  &
\end{align}
Then, only three (adimensional) parameters are required to specify the
solution curves up to similarity transformations. 

In anticipation of later results, we define 
\begin{align}
  X & = \blambda_2 w, &
  Y = \frac{\bc w^2}{\sqrt{1-w^2}},
\end{align}
and consider $(X,Y,p^2)$ as our independent (dimensionless)
parameters. For convenience, we also define the radius $R$ in $(X,Y)$-space, 
\begin{align}
  R^2 = X^2+Y^2.
\end{align}

Curves with $X=0$ correspond to classical elasticae, where the
Lagrange multiplier on torsion vanishes, $\lambda_2=0$. Those solutions
are discussed in detail in \cite{langer-singer-1984}. Curves with
$Y=0$ have constant torsion ($c=0$) and are called elastic rods.

A parity transformation in real space corresponds to a point reflection
of coordinates $(X,Y)\mapsto (-X,-Y)$. Parity invariance implies
symmetry (or anti-symmetry) of all features in the $(X,Y)$ plane under
a $180$-degree rotation, as illustrated in the figures below.

Based on the set of parameters $(X,Y,p^2)$, equation (\ref{eq:link2})
becomes an equation for $w^2$,
\begin{align}
 w^2 = Y^2 + p^2,
\end{align}
showing that the parameter $Y\in [-p',p']$ is bounded, where
$p'=\sqrt{1-p^2}$. Equation (\ref{eq:link1}) becomes an equation for
$\blambda_1$,
\begin{align}
  \blambda_1  &= \frac{1}{2} - \frac{1-R^2}{4w^2} &\text{or} &&
  1-2\blambda_1& = \frac{1-R^2}{2w^2}.
\end{align}
Equation (\ref{eq:link3}) becomes an equation for $\bJ$,
\begin{align}
\bJ =\pm \frac{\sqrt{(1-R^2)^2 + 4(Yw'-Xw)^2}}{4 w^2},
\end{align}
where $w'=\sqrt{1-w^2}$.  Equation (\ref{eq:link4}) becomes an
equation for $\ba$, which we express as 
\begin{align}
  U = w \ba \bJ &= \frac{X(1-R^2) + 2(Yw'-Xw)w}{4\bJ w^2}.
\end{align}
It is convenient to define the conjugate of $U$,
\begin{align}
V  &= \frac{Y(1-R^2) - 2(Yw'-Xw)w'}{4\bJ w^2},
\end{align}
such that $U^2 + V^2 = X^2 + Y^2 = R^2$. 

\begin{figure}[h]
   \includegraphics[width=\linewidth]{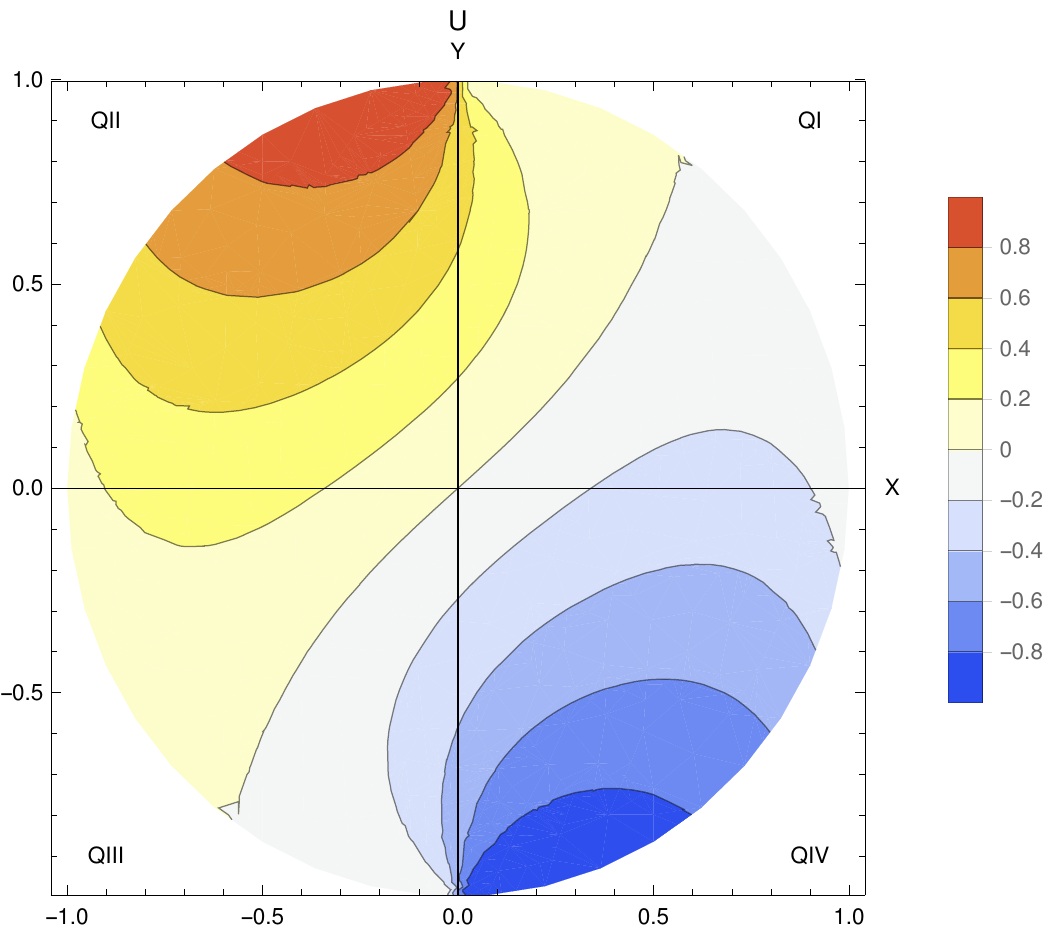}
  \caption{Contours of constant $U$ under the closure condition
    $\Delta z=0$.}
  \label{fig:U}
\end{figure}
\begin{figure}[h]
   \includegraphics[width=\linewidth]{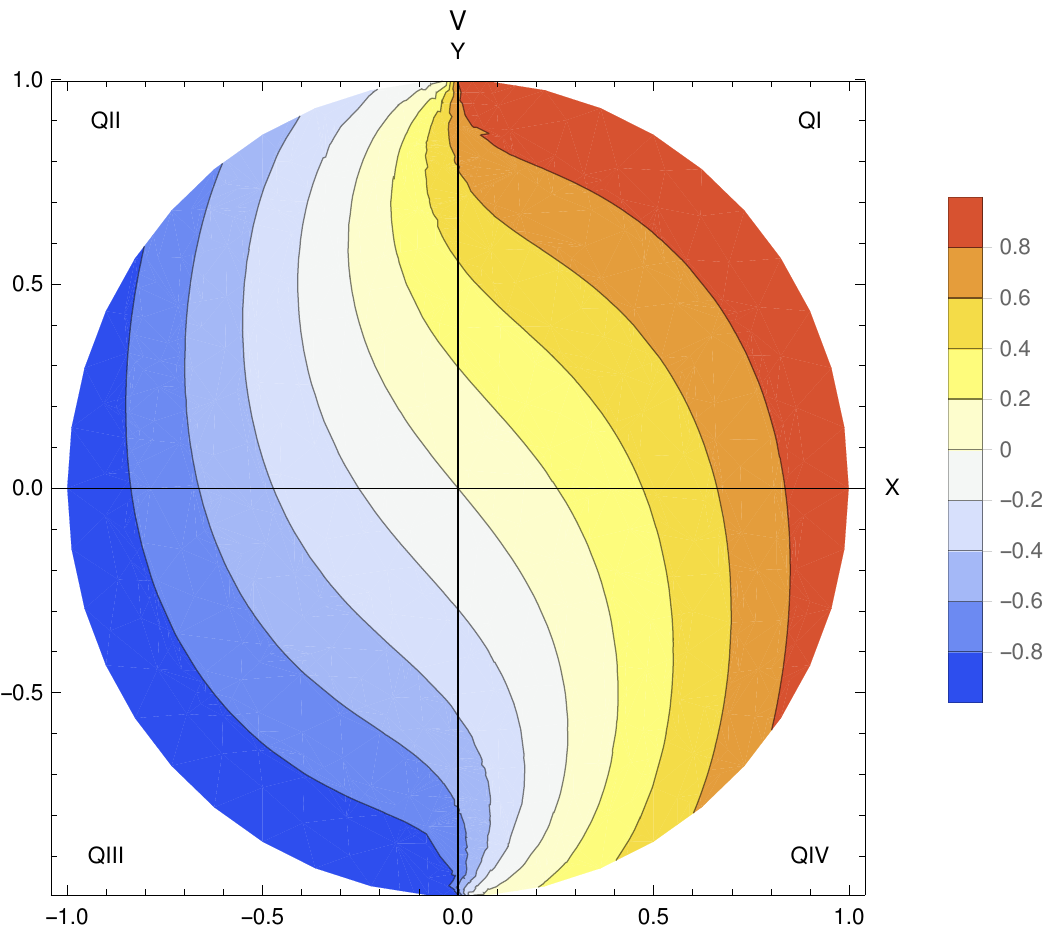}
  \caption{Contours of constant $V$ under the closure condition
    $\Delta z=0$.}
  \label{fig:V}
\end{figure}

\subsection{Cylindrical coordinates}

Exploiting the freedom of choosing the origin and orientation of the
coordinate system (translational and rotational symmetry), we may
align both constants of motion, $\bm{J}= J \hat{z}$ and
$ \bm{A} = a \bm{J} = aJ\hat{z}$ along the vertical axis. We will
assume that $J>0$ defines the vertical direction whereas $a$ can be
positive or negative (the other branch of solutions are obtained by
reversing the arc-length variable $s\mapsto -s$). In this special
reference frame, it is possible to represent the curve analytically
using cylindrical coordinates
$\bm{x}(s)=(r \cos\varphi,r\sin\varphi,z)$, where $r(s)$, $\varphi(s)$
and $z(s)$ are functions of arc-length. Indeed, since
$\bm{x}=r\hat{r} + z\hat{z}$ and $\bm{A}\times\bm{J}=0$ in this
special reference frame, the radial coordinate is immediately deduced
from the square of the vector field $\bm{I}$,
\begin{align}
  I^2 = \kappa^2 + \lambda_2^2 = J^2(r^2+a^2)
   \iff r(s) = \sqrt{\frac{\kappa^2 +\lambda_2^2- a^2J^2}{J^2}}.
\end{align}
Somewhat counter-intuitively, the extrema of the curvature function
and the major radius of the curve occur at the same points. We list:
\begin{align}
  r^2_{max} 
      &=\frac{\kappa_0^2}{J^2} \frac{p^2 + V^2}{w^2}, &
  r^2_{min}
      &=\frac{\kappa_0^2}{J^2}\frac{V^2}{w^2}.
\end{align}
We note that when $V= 0$, $r_{min}= 0$, and
the curve self-intersects on the vertical $z$-axis.

Expressing the tangent vector as
$\bm{t}(s) = \bm{x}' = r'\partial_r\bm{x} +
\varphi'\partial_\varphi\bm{x} + z'\partial_z\bm{x}= r'\hat{r} +
r\varphi'\hat{\varphi} + z'\hat{z}$, one obtains a differential
equation for the vertical coordinate,
\begin{align}
\label{eq:diff_z}
  z' = \bm{t}\cdot \hat{z}
        =  \bm{t}\cdot\frac{\bm{J}}{{J}}
       =  \frac{\kappa^2-2\lambda_1}{2J},
\end{align}
as well as for the toroidal angle,
\begin{align}
  \bm{I}\cdot \bm{t}= \lambda_2 = a J z' - Jr^2\varphi', 
\end{align}
which we express as
\begin{align}\label{eq:diff_phi}
\frac{d  \varphi}{dt} = \frac{awJ}{\kappa_0} \frac{ \kappa^2 -2\lambda_1 - 2\lambda_2 /a}{\kappa^2 + \lambda_2^2 -a^2J^2}
  = U + \frac{N}{1 - M \sn^2(t,p)},
\end{align}
where the following parameters are defined:
\begin{align}
  M & =
 \frac{p^2}{p^2+V^2},\\
  N &= 
U\left(\frac{1-R^2}{2(p^2+V^2)} - 1 \right) -\frac{4w^2\bJ X}{2(p^2+V^2)}.
\end{align}
Analytic solutions of the differential equations (\ref{eq:diff_z}) and
(\ref{eq:diff_phi}) exist in terms of Jacobi theta
functions~\cite{ivey-singer-1999}.

The following useful identity can be proven (see appendix
\ref{sec:proof}):
\begin{align}
  N^2 = \frac{(1-M)(M-p^2)}{M} = V^2(1-p^2-V^2),
\label{eq:important_id}
\end{align}
indicating that $p^2 \leq M \leq 1 \iff 0\leq V^2 \leq 1-p^2$, as well
as the fact that $N=0$ when $V = 0$ and $|V|=p'$ (independent of $X$
and $Y$).

By the property (\ref{eq:killing}), the curvature vector is expressed
in terms of the cylindrical coordinates as
\begin{align}
  \kappa\bm{n} &= \kappa \bm{b}\times\bm{t} = \bm{A}\times\bm{t}+(\bm{x}\times\bm{J})\times\bm{t}\nonumber
\\&=-rJ(a\varphi'+z')\hat{r} + r'J(a\hat{\varphi} + r\hat{z})\label{eq:kappacyl}.
\end{align}
This vector is purely radial at extremal points where
$\kappa'=r'=0$. Evidently, the normal vector is always pointing
radially inwards at the maximum radial position ($t=0$),
\begin{equation}
    \label{eq:kappamaxrad}
    \bm{n}\cdot\hat{r}\big|_{r_{max}} = -1.
\end{equation}
After some tedious and uninteresting algebra starting from
(\ref{eq:kappacyl}), it is found that the direction of the normal
vector at the minimal radial position depends only on the sign of $Y$
and $V$,
\begin{equation}
  \label{eq:kappaminrad}
\bm{n}\cdot\hat{r}\big|_{r_{min}} =  \sgn(V)\sgn(Y),
\end{equation}
where $\sgn(Y)=Y/|Y|$, $\sgn(V)=V/|V|$.

\section{Closure conditions}
\label{sec:closures}
\subsection{Periodicity of curvature}
The curvature, the torsion and the radial coordinate of the solution
curves are periodic functions in $t$ by virtue of $\sn(t + 2K(p),p) = -\sn(t,p)$,
where $K(p)$ is the complete elliptic integral of first
kind. Their period in the arc-length variable $s$ defines the fundamental length $l$,
\begin{align}
\kappa_0 l  = 4  w K(p).
\end{align}
Periodicity of the curvature and torsion functions however does
not guarantee that the curve is closed. The helix is an example of a
non-closing curve where the constant curvature and torsion functions
are trivially periodic. As it will be seen in the following, there are
in fact no non-planar closed solution curves with the same periodicity
as the curvature function. Closed $C^\infty$-curves may
however occur when one lets the parameter $s \in [0, n l]$ extend over
an integer multiple $n\in \mathbb{N}^*$ of the fundamental
length, thereby stringing together identical segments of length $l$, and requiring that i) the vertical displacement vanishes and ii) the toroidal angle returns to the same value modulo $2\pi$. The total
length of the curve is then $ L = n l$ (and the integer $n$ can be
treated as a dependent variable of $X$, $Y$ and $p^2$).

\subsection{Vertical displacement}
\label{sec:zclosure}
By requiring that the net vertical displacement vanishes,
$\int_0^L z'ds = z(L)-z(0) = \Delta z = 0$, the first closure
condition is obtained as
\begin{align}
 < \kappa^2> =   2\lambda_1,
\end{align}
where
\begin{align}
  <\kappa^2> = \frac{1}{L}\int_0^L \kappa^2 ds = \frac{\kappa_0^2}{w^2}\left(w^2+\frac{E(p)}{K(p)}-1\right).
\end{align}
This closure condition does not depend on the size of the curve
(adimensional) and determines the value of $p^2$ as a function of $X$
and $Y$ through
\begin{align}
  X^2 + Y^2= U^2+V^2= R^2 = A(p)  
\end{align}
where
\begin{equation}
A(p) =   2\frac{E(p)}{K(p)} - 1.
\end{equation}
The function $A(p)$ monotonically decreases from $A(0)=1$ to
$A(1)=-1$, passing through zero at
$p_{max}^2=0.8261\ldots$
. The $(X,Y)$ parameters defining curves whose end-points are at the
same height are thus conveniently limited to the unit disk. Although
more stringent bounds exist, it is useful to note that
$A < 1-p^2=p'^2$, $\forall p>0$.

The closure condition implies that \mbox{$z'=(\kappa^2-<\kappa^2>)/2J$} has a
purely oscillatory behaviour, zero mean and the same periodicity as
the curvature function, i.e.~$z(s)$ crosses zero twice per segment
where $r(s)$ is extremal. The vertical coordinate has thus one maximum
and one minimum per segment, alternating between the extrema of the
radial coordinate. The projection on the $(r,z)$ plane is therefore
isotopic to a circle; the solution curves are embedded on tori of
revolution under the first closure condition. Closed curves on this
torus represent a set of measure zero in the parameter space $(X,Y)$.

The edge of the unit disk corresponds to $p\rightarrow 0$,
$K\rightarrow \pi/2$, $w\rightarrow |Y|$, $X^2\rightarrow 1-Y^2$,
$M\rightarrow 0$ and features curves with constant curvature
$\kappa\rightarrow \kappa_0 = 2\pi n |Y|/L$ and constant toroidal variation
$d\varphi/dt\rightarrow U+N$. The torsion function is therefore also
constant and equal to
\begin{equation}
  \tau \overset{p\rightarrow 0}{\longrightarrow}  \frac{2\pi n}{L}|X|\frac{\sgn(X) + \sgn(Y)}{2}
\end{equation}
where $\sgn({X})=X/|X|$. 

At the edge of the unit disk in the first and third quadrants (QI and
QIII), the torsion function $\tau \rightarrow \kappa_0|X|/|Y|$ is
non-zero, yielding helices. The radial excursion of the curves
$r_{max}-r_{min}\rightarrow 0$ however shrinks to zero at the same
time as
$\bJ\rightarrow 0 \Rightarrow r_{min}\kappa_0\rightarrow \infty$. To
respect the finite length condition, the curvature and torsion
function must therefore diverge $\kappa_0,\tau\rightarrow \infty$ (at
fixed ratio). It can be shown that
$|U|\rightarrow 0, |V|\rightarrow 1$ at the edge of QI and QIII, such
that $|N|\rightarrow 0$ and the toroidal excursion vanishes,
$d\varphi/dt\rightarrow 0$. Consequently, the number of segments needs
to diverge $n\rightarrow\infty$ in order to form closed curves. The
limiting solution curves in QI and QIII are somewhat pathological and
uninteresting for the design of stellarators.

At the edge of the unit disk in the second and fourth quadrants (QII
and QIV), the torsion function vanishes, yielding planar circular
curves. One can show that $U\rightarrow Y$,$V\rightarrow X$ and
$N\rightarrow Y$, yielding a finite toroidal displacement per segment
of $\Delta \varphi/ 2\pi n \rightarrow Y$. The solution curves at the
edge of QII and QIV thus trivially close, provided that $Y$ is a
rational number.

\subsection{Toroidal displacement}
\label{sec:toroidal_displacement}
The second closure condition requires that the variation in the
cylindrical angle at the end of one segment be a rational fraction
$m/n$ of $2\pi$ so that, by combining $n$ identical segments, the
curve closes after $m$ revolutions around the vertical axis. The
condition $\Delta \varphi/2\pi n= m/n$ leads to a countably infinite
family of relations between $X$, $Y$ and $p^2$, labelled by the
integer pair $\{(m,n) | |m|<|n|\}$. For the design of the magnetic
axis of stellarators, we are mostly interested in curves that close
after only one revolution around the vertical axis, i.e.
$m=1$. Knotted configurations~\cite{hudson-2014} would correspond to
$m>1$.

The toroidal displacement per segment can be expressed in terms of
elliptic integrals of third kind $\Pi(\alpha^2,p)$ (see appendix
\ref{sec:elliptic}). The expression is valid for any choice of
parameters $(X,Y,p^2)$ and reduces thanks to identity
(\ref{eq:important_id}) to
\begin{align}
  \frac{\Delta \varphi}{2\pi n} &= \frac{1}{2\pi}\int_0^{2K}\frac{d\varphi}{dt} dt = \frac{1}{2}\left[\frac{UK}{\pi/2} + \frac{N}{\pi/2}\Pi(M,p)\right]\nonumber\\
                 &= \frac{1}{2}\left\{\frac{UK}{\pi/2} + \frac{\sgn(N)}{\pi/2}\left[K E(\xi,p') - \left(K-E\right)F(\xi,p')\right]\right\},
\end{align}
where $\sgn(N)=N/|N|$, $F(\xi,p)$ is the incomplete elliptic integral
of first kind, $K=K(p)=F(\pi/2,p)$, $E(\xi,p)$ is the incomplete
elliptic integral of second kind, $E=E(p)=E(\pi/2,p)$ and the $\xi$
angle satisfies
\begin{align}
\cos  \xi  = \sqrt{\frac{p^2(1-M)}{p'^2M}} = \sqrt{\frac{V^2}{1-p^2}}.
\end{align}
This closure condition again does not depend on the size of the curve.

The toroidal displacement per segment $\Delta\varphi/2\pi n$ is a
continuous function away from $N=0$, where either $V$ crosses zero,
$|V|\rightarrow 0$, or $V$ reaches its limiting amplitude,
$|V|\rightarrow p'$. There is no problem in the latter case since
$\xi\rightarrow 0$ and $\Delta \varphi\rightarrow 2nUK$ (under the
first closure condition, this uninterestingly happens only on the edge
of the unit disk in QI and QIII where $|V|\rightarrow 1$ and
$|U|\rightarrow 0$). In the former however, $\xi\rightarrow \pi/2$,
and the Legendre relation,
\begin{align}
K(p) E(p') + E(p)K(p') - K(p)K(p') = \pi/2,
\end{align}
implies that the toroidal displacement per segment tends to
\begin{align}
\label{eq:torsion_discontinuity}
\frac{\Delta \varphi}{2\pi n} \overset{|V|\rightarrow 0}{\longrightarrow}  \frac{1}{2} \left[\frac{UK}{\pi/2} + \sgn(N)\right]  .
\end{align}
The toroidal displacement per segment is thus discontinuous crossing
through the $V=0$ line, where the curves self-intersect at $r=0$. The
jump is exactly $1$.

\begin{figure}[h]
  \centering
   \includegraphics[width=\linewidth]{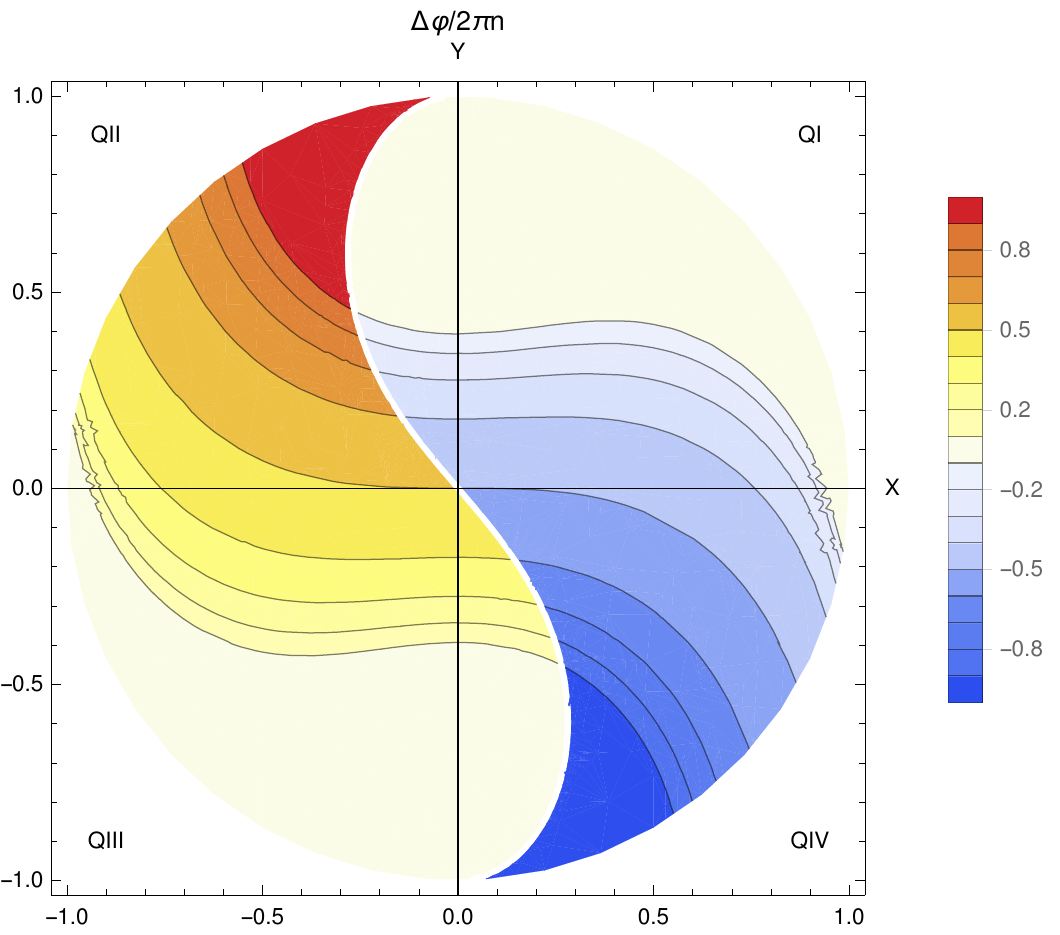}
  \caption{Level curves of the toroidal displacement
    $\Delta \varphi/2\pi n$ under the closure condition $\Delta z=0$
    at the values $\pm 1/2, \pm 1/3,\pm 1/4,\pm 1/5,\pm 1/6$ and at
    the complimentary values $\pm 2/3, \pm 3/4, \pm 4/5, \pm 5/6$.}
  \label{fig:deltaphi}
\end{figure}

Assuming that the first closure condition $\Delta z=0$ is in place, it
turns out that $\sgn(N)=-\sgn(V)$. Figure \ref{fig:deltaphi} displays
the contours of $\Delta\varphi/2\pi n$ at the values
$\pm 1/2, \pm 1/3,\pm 1/4,\pm 1/5,\pm 1/6$ and at the complimentary
values $\pm 2/3, \pm 3/4, \pm 4/5, \pm 5/6$.
It is observed that one can deform a closed $(-m,n)$-curve starting at
$Y=-m/n$ in QIV into a $(n-m,n)$-curve arriving at $Y=1-m/n$ in QII by
following a rational contour of $\Delta\varphi/2\pi n$. In particular,
starting from a once-covered circle on the edge of the unit disk in
the fourth quadrant at $(X,Y)=\sqrt{n^2-1}/n,-1/n)$, the curve can be
deformed into an elastic rod where the contour-line intersect the
$X$-axis ($Y=0$), followed by an elastica where the contour-line
intersects the $Y$-axis ($X=0$), then into a self-intersecting curve
with $n$ lobes where the contour-line crosses the $V=0$ line, then a
$(n-1,n)$ torus knot in the second quadrant where $V<0$ line, and
finally into an $(n-1)$-covered circle (knotted) at the boundary of
the second quadrant at $(X,Y)=(\sqrt{2n-1}/n,1-1/n)$.
The movies in the supplementary material show this sequence for the
$n=3$ and $n=5$ family of closed elastic curves.

\section{Linking, Twisting and Writhing of curves and magnetic fields}
\label{sec:link-twist-writhe}
The average torsion over multiples of the fundamental length can be expressed in terms of elliptic integrals of third kind. The result holds for both open and closed solution curves of
the Euler-Lagrange equations and arbitrary parameters $(X,Y,p^2)$.

\begin{figure}[h]
   \includegraphics[width=\linewidth]{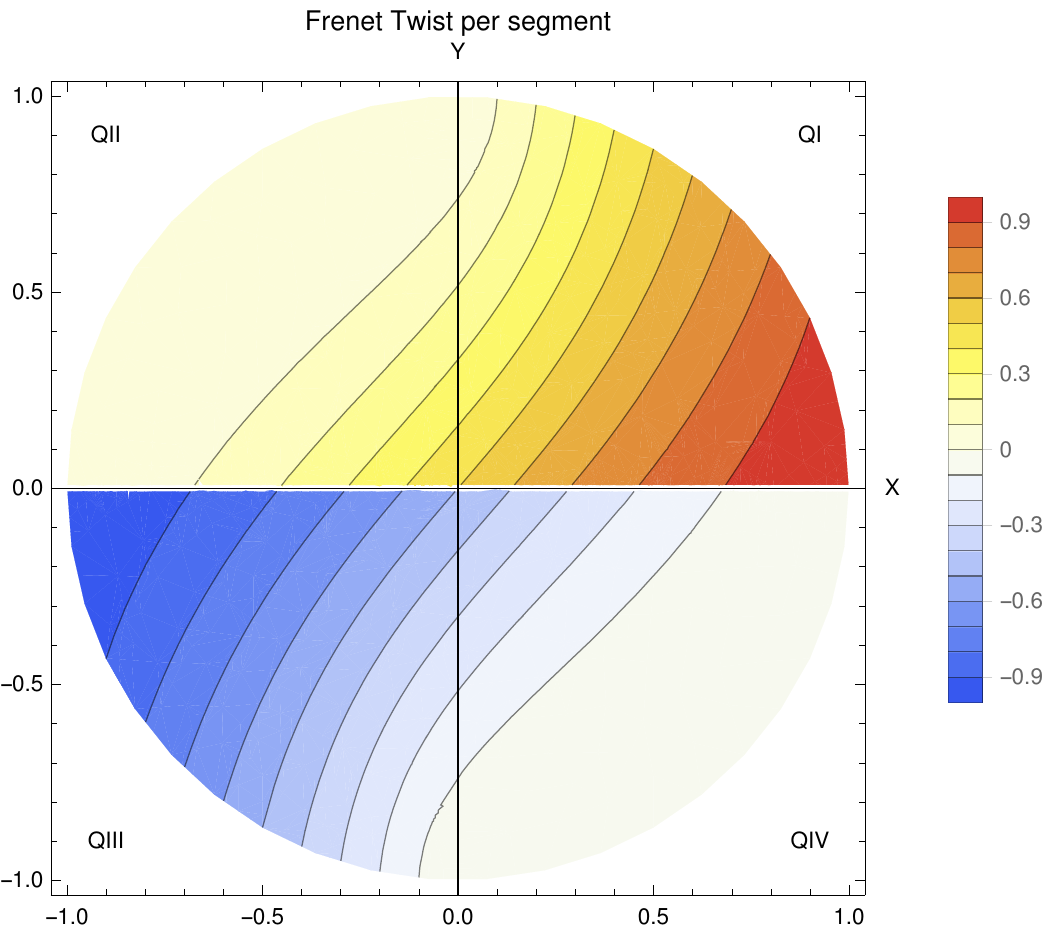}
  \caption{Integrated torsion per segment $\twistfrenet/n$ of a
    solution curve under the closure condition $\Delta z=0$.}
  \label{fig:tauavg}
\end{figure}

In the case where $p^2<w^2$, i.e. $|Y|>0$,
\begin{align}
<  \tau  > &= \frac{1}{L}\int_0^L\!\!\! \tau ds = \frac{1}{K}\int_0^K\!\!\! \tau dt = \frac{\kappa_0}{2}\left[\blambda_2 + \bc\frac{\Pi\left(\frac{p^2}{w^2},p\right)}{ K }\right],
\end{align}
and hence the integrated torsion is
\begin{multline}
\twistfrenet = \frac{n}{2} \left\{\frac{X K}{\pi/2} \right.\\
+\left.\frac{\sgn(Y)}{\pi/2}\left[KE(\chi,p') - \left(K-E\right)F(\chi,p')\right]\right\},
\end{multline}
where the $\chi$ angle satisfies
\begin{align}
  \sin\chi = \frac{w'}{p'}= \sqrt{\frac{1-w^2}{1-p^2}} = \sqrt{1-\frac{Y^2}{1-p^2}}.
\end{align}
Figure \ref{fig:tauavg} shows a contour plot of the integrated torsion
per segment $\twistfrenet/n$ on the $(X,Y)$ unit disk where the
closure condition $\Delta z = 0$ is imposed to define the value of
$p^2$ (see section \ref{sec:zclosure}).

As $|Y|\rightarrow 0$, $p^2\rightarrow w^2$ and
$\chi\rightarrow \pi/2$, the Legendre relation implies that the
integrated torsion tends to
\begin{align}
\label{eq:torsion_discontinuity}
\twistfrenet \overset{|Y|\rightarrow 0}{\longrightarrow}   \frac{n}{2} \left[\frac{XK}{\pi/2} + \sgn(Y)\right]  .
\end{align}
and is discontinuous as $Y$ changes sign. The jump is exactly $
n$. For the special case of elastic rods, $p^2=w^2 \iff c = Y=0$, the
constant torsion function is held at the intermediate value
\begin{align}
\tau\big|_{Y=0} = \frac{\lambda_2}{2} 
= \frac{n}{2}\frac{X K}{\pi/2}\frac{2\pi}{L}.
\end{align}
In this case, the curvature function touches zero (and its derivative
is discontinuous) at exactly $n$ \emph{inflexion} points along the
curve. Although the curve itself is smooth and well-defined, its
representation via the Frenet frame breaks down. It is tempting to
believe, in light of Mercier's formula~\cite{mercier-1964}, that the
discontinuity of the integrated torsion in equation
(\ref{eq:torsion_discontinuity}) has a physical impact on the on-axis
rotational transform of a magnetic field, but this would be ignoring
the role of the \emph{integer of topological
  origin}~\cite{helander-2012}. This number actually corresponds to
the linking of the Frenet frame, $\linkingfrenet$, i.e. how many times
the normal vector wraps around the curve. It is an integer
$\linkingfrenet = Z\in \mathbb{Z} $ for closed periodic curves and a
topological invariant of the Frenet framing.

In~\cite{ricca-moffatt,moffatt-ricca}, it is demonstrated by
continuously deforming a generic curve through states of inflexion
that both the integrated torsion $\twistfrenet$ and the linking of the
Frenet frame $\linkingfrenet$ are discontinuous functions. Fortunately
however, the jump has to be exactly the same for both functions such
that their difference, the so-called Writhe,
\begin{equation}
\label{eq:calugareanu-frenet}
\linkingfrenet - \twistfrenet = Z - \tfrac{L}{2\pi}<\tau> = \writhefrenet,
\end{equation}
is a smoothly varying function across a state of inflexion. This is a
corollary of the well-celebrated \Calugareanu
theorem~\cite{calugareanu-1959} applied to the Frenet frame. The
Writhe is a geometric property of the curve, which is well-defined in
the presence of inflexion points or straight segments. It represents
the average crossing number over all projections and can be evaluated
via a double Gauss integral~\cite{fuller-1971}. Most importantly, the
Writhe is independent of the framing of the curve, i.e.
$\writhefrenet = \writhe$.

The \Calugareanu theorem acquires a special interpretation when used
to quantify how many times a neighbouring field-line links to the
magnetic axis,
\begin{equation}
\label{eq:revisit_mercier}
  \linking = \writhe + \twist = \linkingfrenet - \twistfrenet + \twist,
\end{equation}
i.e. the field-line linking (not necessarily integer) is equal to the
linking of the Frenet frame \textbf{minus} the integrated torsion of
the axis, plus the Twist $\twist$ of the magnetic field. The Twist of
the magnetic field can be shown to originate independently and almost
exclusively from parallel current (curl) and the rotation of the
elliptical boundary. This statement will be further detailed in a
future publication. The important message is that the physically
relevant quantity for the on-axis rotational transform in relation to
non-planar geometry is the \textbf{Writhe} of the magnetic axis.

A standard definition of the rotational transform is the average
linking of neighbouring field-lines per toroidal revolution,
i.e. $\linking$ divided by the total toroidal displacement of the
curve,
\begin{equation}
  \label{eq:iota}
 \frac{\iota}{2\pi} := \frac{\linking}{\Delta\varphi} \sim \frac{\writhe}{\Delta\varphi}.
\end{equation}
The presence of the integrated torsion in Mercier's formula appears to
be, in light of equation (\ref{eq:revisit_mercier}), an artefact of
the Frenet frame. Expression (\ref{eq:iota}) is noted to be suitable
for vacuum fields (no parallel currents) surrounded by a circular
boundary where the Twist of the magnetic field can be neglected.

To work out the Writhe of the unknotted closed solution curves along
the level contour $\Delta\varphi/2\pi n=-1/n$, we start at the edge of
QIV at $(X,Y)=(\sqrt{n^2-1}/n,-1/n)$ where the curve is composed of
$n$ planar arc-lengths forming a (once-covered) circle. Since $Y<0$
and $V>0$ along $\Delta \varphi/2\pi n=-1/n$, the normal vector never
points outwards (see equation \ref{eq:kappamaxrad} and
\ref{eq:kappaminrad}) such that the neighbouring curve
$\bm{x}(s)+\epsilon \bm{n}(s)$ easily detaches from $\bm{x}(s)$,
$\forall \epsilon$. This implies that the Frenet frame is not linking,
and the integer $Z=0$. The Writhe $\writhe = -\twistfrenet$ is simply
minus the integrated torsion along the level contour
$\Delta \varphi/2\pi n=-1/n$ where $Y<0$, and for these solution
curves, the rotational transform matches the integrated torsion
identically,
\begin{equation}
  \label{eq:iotazero}
  \iota^0 = \twistfrenet.
\end{equation}
The integrated torsion is negative in QIV, vanishing at the edge of
the unit disk. The subsequent vanishing of the Writhe is consistent
with the fact that, for planar curves, the Writhe is equal to the
number of crossings, and the once-covered circle is devoid of them.

Passing through the $Y=0$ line into QI, the solution curves
experiences $n$ inflexion points where the curvature vector
vanishes. Because the integrated torsion jumps by $+n$ but the Writhe
must remain continuous, we deduce that the Frenet frame must link by
the same amount, i.e. $Z=n$. Indeed, when $Y>0$ along
$\Delta\varphi/2\pi n=-1/n$, the normal vector points radially inwards
at $r_{max}$ but outwards at $r_{min}$. Knowing that the solution
curve lies on a torus of revolution, the neighbouring curve
$\bm{x}(s)+\epsilon \bm{n}(s)$ behaves as the torus' centre line, and
is being linked once per periodicity by $\bm{x}(s)$. The Writhe of
closed solution curves is thus $\writhe = n - \twistfrenet$ along
$\Delta\varphi/2\pi n=-1/n$ where $Y>0$. The rotational transform is
then given by
\begin{equation}
  \label{eq:iotaplus}
\iota^+ = -n\left(1 - \frac{\twistfrenet}{n}\right).
\end{equation}

The level contour $\Delta\varphi/2\pi n=-1/n$ ends in QII on the $V=0$
line, where closed solution curves self-intersect at the origin of the
cylindrical axis. The level contour however connects with that of
$\Delta\varphi/2\pi n=(n-1)/n$ in the region where $V<0$, yielding a
continuous deformation (homotopy) from unknotted solutions to
$(n-1,n)$-torus knots in QII. As a particular case of the general
formula $C=\min[(p-1)q,(q-1)p]$ for the number of crossings of
$(p,q)$-torus knots, the conjugate solution curves feature
$C=(n-2)n$ crossings (all negative). They tend to
$(n-1)$-covered flat circles at the edge of the unit disk in QII,
where the torsion vanishes. Consequently, the value of the Writhe
approaches $-C$ and we deduce that the Frenet linking is
$ Z = -C= n(2-n)$ along the entire level contour
$\Delta \varphi/2\pi n = (n-1)/n$. In QII where $V<0$, the Writhe is
thus $\writhe = n(2-n) - \twistfrenet$ and the rotational transform
(\ref{eq:iota}) reads
\begin{equation}
  \label{eq:iotaminus}
  \iota^- = -n\left(\frac{n-2}{n-1} + \frac{\twistfrenet}{n(n-1)}\right).
\end{equation}
Expression (\ref{eq:iotaminus}) has the disadvantage that $\iota^-$
tends to the finite value $n(n-2)/(n-1)$ at the edge of QII, where the
knot becomes a $(n-1)$-covered circle. This conflicts with the idea
that rotational transform measures on a Poincaré plot the poloidal
increment $\Delta \theta$ of neighbouring field-lines around the
magnetic axis, $\iota / 2\pi = \Delta \theta /\Delta\varphi$;
field-lines that follow the Frenet frame are fixed points on a
Poincaré plot and do not progress poloidally, i.e.
$\Delta \theta \propto \iota \rightarrow 0$ is expected to vanish at
the edge of QII. For knotted configurations, it thus seems logical to
offset the definition of rotational transform by the linking of the
curve with respect to a fixed axis, e.g. the vertical axis. The latter
choice is known as the \emph{blackboard framing} and yields the
so-called \emph{Kaufmann self-linking number} $\linkingself$
\cite{pohl-1968}. The effective rotational transform per curve
periodicity is now defined as
\begin{equation}
  \label{eq:effective_iota}
  \iota_{\text{eff}} := \linking - \linkingself \sim \writhe - \linkingself
\end{equation}
The prescription (\ref{eq:effective_iota}) suits unknotted
configurations, where $\linkingself = 0$, so that results
(\ref{eq:iotazero}) and (\ref{eq:iotaplus}) remain unchanged.  The
Frenet and blackboard framings coincide at the edge of the unit disk,
so that for knotted solution curves $\linkingself=\linkingfrenet=\mp C$
and we obtain
\begin{equation}
\iota_{\text{eff}}^-= -\twistfrenet.  
\end{equation}

\begin{figure}
   \includegraphics[width=\linewidth]{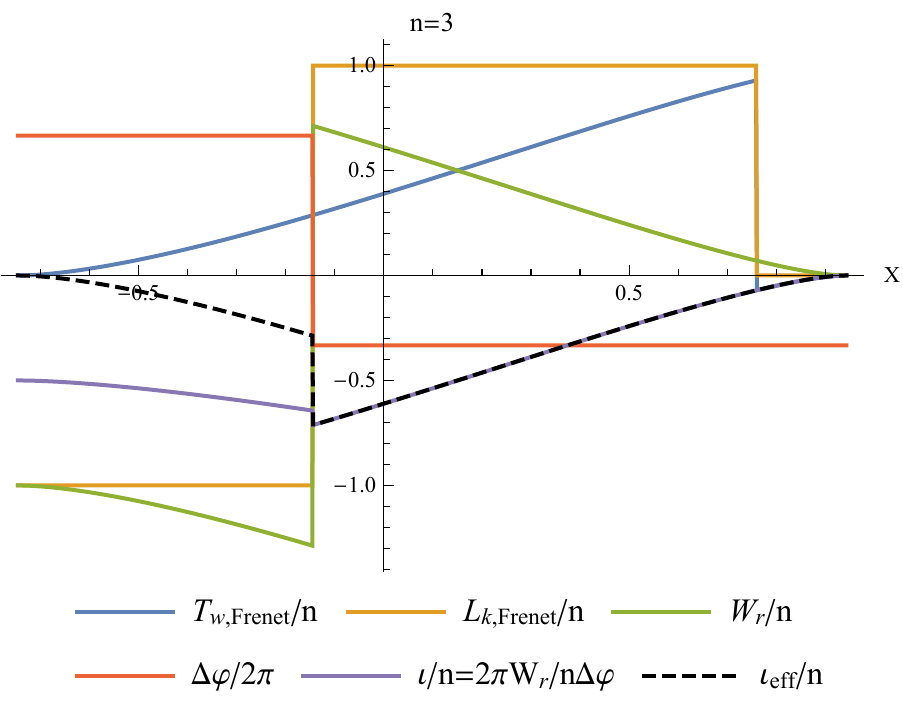}
  \caption{Behaviour of the integrated torsion (blue), Frenet linking
    (orange), Writhe (green), toroidal displacement (red), on-axis
    rotational transform (purple) and effective rotational transform
    (dashed black) per segment for the $n=3$ family of closed solution
    curves as a function of the parameter $X$. The two discontinuities
    correspond to the crossing of the $Y=0$ line (on the right) and
    the $V=0$ line (on the left).}
  \label{fig:family3}
\end{figure}

The integrated torsion $\twistfrenet$ is continuous across the $V=0$
line, as seen on figure \ref{fig:tauavg}, so the jump $n(n-1)$ in the
Frenet linking must therefore compensate the jump in the Writhe, again
by the \Calugareanu theorem. Figure \ref{fig:family3} summarises the
behaviour of the integrated torsion per segment $\twistfrenet/n$, the
Frenet linking per segment $\linkingfrenet/n$, the Writhe per segment
$\writhe/n$, the toroidal displacement per segment
$\Delta\varphi/2\pi n$ and the on-axis rotational transform per
segment $\iota/n$ and $\iota_{\text{eff}}/n$ as a function of the
parameter $X$ while deforming the $n=3$ family of solution curves
along level contours $\Delta \varphi/2\pi n=-1/3$ and
$\Delta\varphi/2\pi n=2/3$. The evolution is also illustrated in the
supplementary material.

\begin{figure}
  \includegraphics[width=\linewidth]{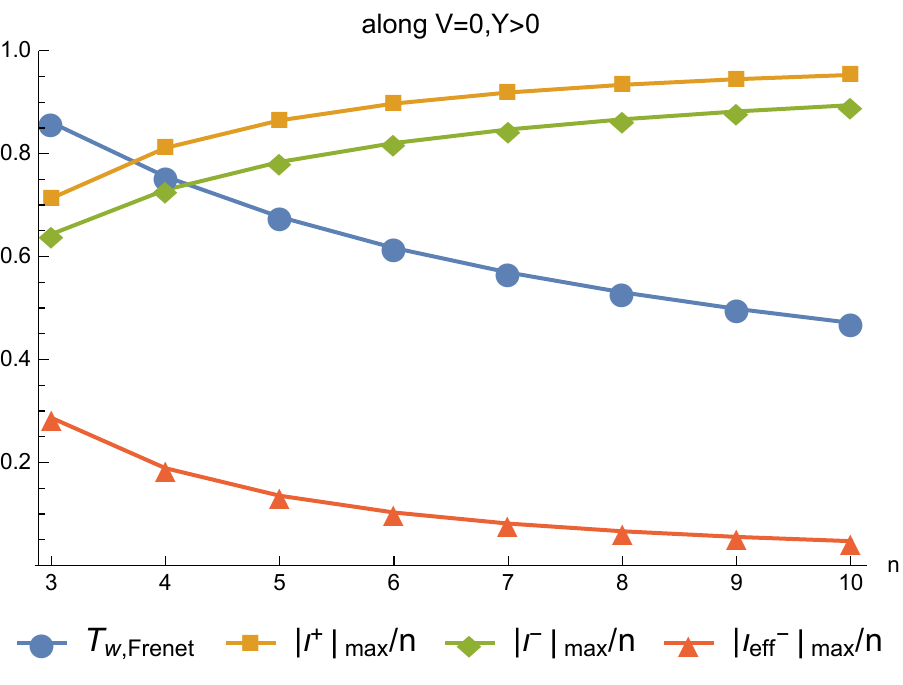}
  \caption{(blue) Integrated torsion $\twistfrenet$ along the $V=0$ as
    a function of the number of segments $n$. (orange) maximum
    achievable on-axis rotational transform per segment of unknotted
    configurations, (green) of the conjugate $(n-1,n)$-torus knots,
    (red) maximum effective rotational transform of $(n-1,n)$-torus
    knots per segment.}
  \label{fig:iotamax}
\end{figure}

According to formulae (\ref{eq:iotazero}-\ref{eq:effective_iota}), the
rotational transform (purple curve on figure \ref{fig:family3})
reaches its extreme value at the $V=0$ line for any value of $n$. As
shown on figure \ref{fig:iotamax}, the integrated torsion
$\twistfrenet|_{V=0}<1$ is always smaller than unity on the $V=0$ line
and decreases with the number of segments. Consequently, the on-axis
rotational transform
$|\iota^+|_{max} > |\iota^-|_{max}>|\iota^-_{\text{eff}}|_{max}$ is
greater in absolute value for unknotted configurations than for knotted
ones, as highlighted by the orange, green and red markers on figure
\ref{fig:iotamax}.


\section{Conclusion}
\label{sec:conclusion}
The benign question of finding closed curves with minimal bending
energy but fixed integrated torsion led to the study of a rich class
of elasticae. Several interesting discontinuities appear explicitly in
the analytic solutions, which were given in terms of elliptic integral
of third kind. The discontinuity of the integrated torsion, where the
solution curves pass through a state of inflexion, is understood as an
artefact of the Frenet frame. The discontinuity of the toroidal
displacement per segment, where the solution curves self-intersect
through the origin, is understood as an artefact of the cylindrical
coordinate system. Thanks to the \Calugareanu theorem, one of the
three relevant quantities, namely Writhe when crossing the $Y=0$ and
Twist when crossing the $V=0$ line, remained continuous, thereby
imposing the jumps in the other two to mutually cancel. Although
everything could have been calculated explicitly, the strong
constraint that is the \Calugareanu theorem helped deduce the Linking
of the Frenet frame for all closed solution curves as well as quantify
the rotational transform purely arising from the non-planar geometry
of the magnetic axis.

Despite the obvious scaling of the on-axis rotational transform with
the number of repeated segments $n$, it is found that the extremum of $\iota/n$
is achieved by unknotted magnetic axes that are about to
self-intersect. Whether these solutions can be realised by the vacuum
fields of a coil design remains to be addressed. \amend{The detailed
  study of these optimal curves nevertheless highlights the leading
  role of the \emph{Writhe} component. This fact is not apparent from
  Mercier's formula but important to interpret figure \ref{fig:tauavg},
  which deceptively shows that the maximum integrated torsion is
  achieved by a magnetic axis with constant torsion. The beneficial
  effect of Writhe suggests the possibility of designing plasmas with
  5-10 times the on-axis rotational transform of current
  stellarators. The usual Fourier representation is unsuited to access
  such configurations through numerical optimisation since the toroidal angle will generally be a
  non-monotonic function of the axis' arc-length.}

We conclude by noting that such optimal curves seem to occur in other
natural phenomena such as the super-coiling of DNA~\cite{dna}, the
vortices of under-water air bubbles~\cite{kleckner-2013}, flux-tubes
in the solar corona~\cite{torok-2014}, etc... Results herein may be
directly applicable to the interplay between Twist, Writhe and Linking
in those systems.

\section*{Supplementary material}
\amend{The supplementary material contains two movies illustrating the
  continuous deformation discussed in \ref{sec:toroidal_displacement}
  of the $n=3$ and $n=5$ family of closed elasticae (in black) and the
  linking of the Fernet frame in the form of a curve (in red)
  displaced in normal direction.}

\section*{Acknowledgment}
The authors would like to thank stimulating discussions related to
this work with E.Hirvijoki, D.E.Ruiz, A.Brizard, N.Fowkes and
S.I.Abarzhi.

\bibliographystyle{apsrev4-1}
\bibliography{biblio}

\appendix
\section{Proof of identity (\ref{eq:important_id})}
\label{sec:proof}
Express $\bJ$ as a function of $\blambda_1$, $\blambda_2$ and $\bc$
using equation (\ref{eq:link3})
\begin{align}
\bJ^2 &= \frac{1}{4}[(1-2\blambda_1)^2 + (c-\blambda_2)^2]
\end{align}
and $\ba$ from equation (\ref{eq:link4})
\begin{align}
\ba &= \frac{2(c-2\blambda_1\blambda_2)}{(1-2\blambda_1)^2 + (c-\blambda_2)^2}
\end{align}
From (\ref{eq:link1}), express $p^2$ as a function of $\blambda_1$,
$\blambda_2$ and $w^2$
\begin{align}
  p^2 = w^2[1+\blambda_2^2+2(1-2\blambda_1)] -1
\end{align}
and then from (\ref{eq:link2})
\begin{align}
\label{eq:proof}
  w^4(c^2+4\blambda_1-\blambda_2^2-2) - w^2(4\blambda_1-\blambda_2^2-3)-1 = 0
\end{align}
Then
\begin{align}
N^2 -\frac{(1-M)(M-p^2)}{M}
= \frac{4\bJ^2\times\text{eq.}(\ref{eq:proof})}{w^2(1-2\blambda_1 -c \blambda_2 + \blambda_2^2)^2}
\end{align}

\section{Elliptic integral of third kind}
\label{sec:elliptic}

According to \cite[413.01]{byrd-friedman}, the complete elliptic
integrals of third kind can be expressed in terms of incomplete
elliptic integral of first and second kind. For the case II
(circular), i.e. $p^2 < \alpha^2 < 1$, one has
\begin{align}
\Pi(\alpha^2,p) &= \int_0^K \frac{dt}{1-\alpha^2 \sn^2(t,p)} \\
&=  \frac{\alpha \left[ (E-K) F(\xi,p') + K E(\xi,p')\right]}{\sqrt{(\alpha^2-p^2)(1-\alpha^2)}}
\end{align}
where 
\begin{align}
  \sin \xi &= \sqrt{\frac{\alpha^2-p^2}{\alpha^2p'^2}} &  
  p'^2 = 1-p^2
\end{align}

In particular,
\begin{align}
  \Pi(p^2,p) = \frac{E}{1-p^2}
\end{align}
and
\begin{align}
  \Pi\left(\frac{p^2}{w^2},p\right) = \frac{w[(E-K) F(\xi,p') + K E(\xi,p')]}{\sqrt{(1-w^2)(w^2-p^2)}}
\end{align}
for 
\begin{align}
  \sin\xi = \sqrt{\frac{1-w^2}{1-p^2}}  
\end{align}



\end{document}